\documentstyle[12pt]{article}
 
\begin{document}


\begin{flushright}
{\bf 30 November, 1996}
\end{flushright}

\centerline{\large\bf Momentum Resolution Studies for a}
\centerline{\large\bf TPC-Based Central Tracker}

\vskip 2.0 cm

\centerline{B. J. King}

\vskip 0.5 cm

\centerline{DESY}
\centerline{F-OPAL}
\centerline{Notkestrasse 85}
\centerline{22607 Hamburg}
\centerline{Deutschland}
\vskip 0.5 cm
\centerline{email: KING$@$OPL01.DESY.DE}
\vskip 1.5 cm

\centerline{\large\bf Abstract}
   This note describes a study on the momentum resolution of the
TPC-based central tracker design which will be included in the
ECFA/DESY conceptual design report (CDR) for the proposed DESY 500 GeV
electron-positron linear collider. The study found it to be likely
that the central tracker could achieve a
momentum resolution of approximately $10^{-4}\; ({\rm GeV/c})^{-1}$,
particularly if a layer of silicon microstrip detectors were to be
added just inside the TPC.

\vskip 0.5 cm

\pagebreak

\section{Introduction}

   This note describes a study on the momentum resolution of the
TPC-based central tracker design (CT) which will be included in the
ECFA/DESY conceptual design report (CDR) for the proposed DESY 500 GeV
electron-positron linear collider.

   The study had two goals:
\begin{itemize}
  \item To estimate the CT momentum resolution as a function of the
        Time Projection Chamber (TPC) coordinate resolution and level
        of relative misalignment between the TPC and and vertex detector (VD).
  \item To try to characterise the improvements which would be
        expected from adding a layer of silicon microstrip detectors
        (SI) outside the vertex detector and just inside the TPC.
\end{itemize}

   Only the physically important high transverse momentum limit for
the CT momentum resolution is considered:
\begin{equation}
p_t = \infty\; \Leftrightarrow \; \frac{1}{p_t} = 0,
\end{equation}
and only for the barrel region of the CT.

   Section 2 of this paper describes the CT model used for these
studies, which is roughly the model which emerged from the discussions
at the Munich ECFA/DESY workshop~\cite{munich}. For this reason, the
model will be referred to as the ``Munich CT''.

   In section 3, details are given on the method used to determine
the momentum resolution of the CT. Section 4 describes how the
coordinate resolution of the TPC was estimated, based on both the
discussion at the Munich workshop and the known performance of
the ALEPH TPC~\cite{alephtpc}. The performance of the ALEPH central
tracker was also used to obtain an estimate for a reasonable level
of TPC misalignment, as discussed in section 5.

   The results of this study are presented in section 6, consisting of
the momentum resolutions obtained for different alignment and
resolution assumptions, and using different combinations of the
central tracking subdetectors, and conclusions for the study
are given in the final section of the report.

%
%
%

\section{Central Tracker Model}

   Based on the discussions at the Munich ECFA/DESY
workshop~\cite{munich},
the central tracker is assumed to consist of 4 layers of
vertex detector, at radii of 2.5, 3.5, 7.5 and 10 cm, and a TPC
with an active region between 40 cm and 150 cm radius.
It is expected that the general conclusions of this study will
apply for other similar configurations.

   Optionally, a layer of silicon microstrips (SI) is added
at a radius of 32 cm. This is 8 cm inside the inner active radius
of the TPC, with the 8 cm gap corresponding roughly to
the size of the inactive region on the inside of the ALEPH
TPC. One part of the study investigates the effect of increasing
the SI radius to 42 or 52 cm, with the inner radius of the TPC
increased to 50 cm or 60 cm, respectively, to retain the 8 cm
gap.

   The study considers only the barrel region of the CT, which
is defined by the outer corner of the TPC, at $r_{corner}=150$ cm
and $z_{corner}=250$ cm. The corresponding polar angle is
$\arctan(0.6)\; =\; 31$ degrees, which subtends 85.8 percent of the solid
angle. The point resolutions of the vertex detector and silicon layer,
in the coordinate for the bend plane, should be relatively constant
over this range. This should also be fairly true
for the TPC, although the details depend on which effect limits
the TPC resolution -- as discussed in section 4.

   It is easily seen that the area of Si needed to cover the barrel
region is given by:
\begin{equation}
{\rm Si\; area} =  2\times \pi\times r\times (2\times Z_{end}),
\end{equation}
with
\begin{equation}
Z_{end} = z_{corner}/r_{corner} \times r = 1.67 \times r
\end{equation}
giving
\begin{equation}
{\rm Si\; area} = 20.9 \times r^2 \\
\end{equation}
For $r=32$ cm, 2.14 square metres of silicon microstrips are
required. This would cost about 1.3 million CHF at a conservative
total cost per unit area
-- detectors plus readout -- of 60 ${\rm CHF/cm^2}$~\cite{bjkallsi}.

   The vertexing layers and the Si were both assumed to have point
resolutions of 8 microns. The TPC readout is assumed to be radially
segmented into $N_{TPC}$ equal-length pads, each with position
resolution $\sigma_{pad}$. The choice of values for $N_{TPC}$ and
$\sigma_{pad}$ is discussed in section 4.

   In the interests of simplicity, the form of the TPC misalignment
was assumed to be a simple transverse displacement relative to the
SI layer, with a gaussian uncertainty. The magnitude chosen for this
uncertainty is discussed in section 5. Much smaller transverse
displacements of 5 microns were included to describe the misalignment
of SI relative to VD, and VD relative to the beam spot.

\section{Method for Determining Momentum Resolutions}

   This section describes the method used for determining the various
momentum resolutions for the CT, which involves the Monte Carlo-based
generation of large numbers of simulated infinite-momentum tracks.
It begins by giving a derivation of the method for measuring
high momentum tracks using a quadratic fit, followed
by a description of the Monte Carlo-based method for determining
the tracker's momentum resolution.

   In a quadratic approximation, the measured coordinates in the bend
plane of a high momentum track, $f_i(r_i)$, at
the radial positions, $r_i,\;i=1,N$, of the $N$ detector elements,
can be fitted to a "Chebychev-type" quadratic equation:
\begin{equation}
f(r) = a + b(r-m) + 0.25\times c\times ( 2(r-m)^2 - m^2 ),  \label{eq:quad}
\end{equation}
where $m$ is the half-radius of the TPC and the fit is for the
3 parameters, $a,b$ and $c$.

   The relevant parameter for momentum resolution is the coefficient,
$c$, of the quadratic term. For each detector configuration, the
momentum resolution is obtained from
the r.m.s. uncertainty, $\sigma_c$, of a Gaussian fit to the
distribution of $c$ values for 1000 infinite momentum tracks, as follows.

   Using the Lorentz force equation, it is easy to translate the fitted
value of $c$ into a measurement of the transverse momentum,
$p_t$~\cite{pdg}:
\begin{equation}
c = \frac{d^2f}{dr^2} = \frac{B}{0.3\times p_t},            \label{eq:candpt}
\end{equation}
where $f$ and $r$ are in metres, $B$ is in Tesla, $p_t$ is in GeV/c and
the factor of 0.3 accounts for the momentum units and the speed of
light:  $0.3 = 3.10^8\;{\rm m/s} \times  10^{-9} {\rm GeV/eV}$.

   As a clarifying remark about the high momentum limit,
equation~\ref{eq:candpt} shows that the directly measured quantity is
actually the inverse of the track momentum,
$\frac{1}{p_t}$, and it is this quantity which has an approximately
gaussian resolution in the infinite transverse momentum limit.
In practice, the momentum
itself will also have an approximately gaussian distribution for all
the tracks encountered in an experiment, since these tracks  will always
have a low enough momentum that the inverse momentum can be measured
with a fractional uncertainty much less than one. In this case, the
sigmas of the gaussian widths are related by:
\begin{equation}
\sigma(\frac{1}{p_t}) = \frac{\sigma_{p_t}}{p_t^2}.
\end{equation}

   Using this relation,
equation~\ref{eq:candpt} can be converted to an equation for the transverse
momentum resolution,
$\sigma_{p_t}$, in terms of the width, $\sigma_c$ of $c$:
\begin{equation}
\frac{\sigma_{p_t}}{p_t^2} = \frac{0.3 \times  \sigma_c}{B},
                                          \label{eq:ctop}
\end{equation}
in units of $({\rm GeV/c})^{-1}$, where $\sigma_c$ is in units of
$1/{\rm m}$ and $B$ is in Tesla.
This is the basic equation used in the procedure for determining the
tracker resolution, which will now be described.

   The procedure simulates the measurement of infinite-momentum
tracks, which will have zero displacement, $f$, from a
straight line at all radii ($r$):
\begin{equation}
f(r) \equiv 0,\; {\rm at\;all}\;r.
\end{equation}
However, the measured points, at $f(r_i),\; r_i=1,N$, will
differ from zero due to point resolutions and misalignments.

   For each of 1000 simulated tracks, the measured displacements at each
of the $N$ measurement positions were generated randomly in two
stages: (i) independent displacements were generated corresponding
to the point resolutions, and (ii) the misalignments of the
subdetectors were generated and then added to each measurement
point in the subdetector.

   Next, each of the tracks was fitted, using MINUIT, to the
parameterization of equation~\ref{eq:quad}. As expected, the fitted
values of the coefficient of the quadratic term, $c$, formed a
gaussian distribution centred on zero. The resolution parameter,
$\sigma_c$, was obtained from a gaussian fit to this distribution,
and converted to a momentum resolution, $\frac{\sigma_{p_t}}{p_t^2}$,
using equation~\ref{eq:ctop}.

    As a detail regarding the fitting method, the misalignments
will introduce correlations between the coordinate measurements,
$f_i$, so the optimal fit procedure would require the use of an
$N \times N$ covariance matrix. This was regarded as being unnecessarily
complicated, given that the assumed form of the misalignments is,
anyway, unrealistically simple. Instead, correlations between
the measurements were ignored in the fit, but the TPC point
resolutions assumed for the fit were adjusted to compensate
for this. The fit used:
\begin{equation}
\sigma_{fit}^2 = \sigma_{point}^2 + k.\sigma_{align}^2
\end{equation}
where it can be seen that the misalignment variance is added to
the point variance weighted by an adjustable parameter, $k$.
Several fits were performed on the ensemble of 1000 tracks --
one each for different values of $k$ -- and the most precise
value for the momentum resolution was chosen.

\section{Estimation of TPC Point Resolutions}

   The TPC point resolution was estimated from the TPC-alone momentum
resolution, $2.3\times 10^{-4}\;({\rm GeV/c})^{-1}$, that was
quoted~\cite{settles} at the Munich ECFA/DESY woorkshop, for a TPC
with an inner (outer) radius of 40 cm (150 cm) and a 3 Tesla magnetic
field. The performance of the ALEPH TPC~\cite{alephtpc} was
also considered. Before
determining this point resolution, this section first discusses
how the point resolution is parameterized, and then discusses the
physical effects which determine the point resolution of a TPC.

   For the purposes of this study, it is reasonable to assume that
the measurements from the TPC pads are
statistically independent and that $\sigma_{pad}$ scales statistically
according to the number of primary ionization electrons producing
the pad signal. That is, it reduces as the inverse square root of the
pad's radial length. In this approximation, it is possible to define a TPC
position resolution:
\begin{equation}
\sigma_{TPC} \equiv \sigma_{pad} \times L_{pad},
                                               \label{eq:sigmatpc}
\end{equation}
for $L_{pad}$ the pad length and
with units of ${\rm \mu m.cm^{1/2}}$, which is
independent of the radial segmentation of the pads.

   Further, when equation~\ref{eq:sigmatpc} holds, the momentum
resolution for high momentum tracks is also almost independent of
the radial segmentation. The well-known Glueckstein formula~\cite{pdg}
predicts the momentum resolution for a singly charged particle:
\begin{equation}
\sigma(\frac{1}{p_t}) = \frac{1}{0.3 \times B({\rm Tesla})} \times
\frac{\sigma_{pad}}{L^2} \times \sqrt{\frac{720}{N_{TPC}+4}},
                                         \label{eq:gluck}
\end{equation}
with $L$ the radial extent of the TPC, and all units of length in
metres. If it is assumed that the pads occupy the entire radius of the TPC,
\begin{equation}
L = N_{TPC} \times L_{pad},
\end{equation}
and
\begin{equation}
N_{TPC} \gg 4,    \label{eq:nggfour}
\end{equation}
then equations~\ref{eq:sigmatpc} and~\ref{eq:gluck} can be
combined into:
\begin{equation}
\sigma(\frac{1}{p_t}) \simeq \frac{\sqrt{720}}{0.3 \times B({\rm Tesla})} \times
\frac{\sigma_{TPC}}{L^{2.5}},      \label{eq:resapprox}
\end{equation}
which is seen to be independent of the pad segmentation, $N_{pad}$.

   It should also be noted that this argument
would break down for tracks crossing the pads at large angles, for
which the finite radial segmentation would contribute to the pad measurement
uncertainty, $\sigma_{pad}$. This is not a problem for the studies in
this report: in the high momentum limit that we are considering the
track directions are always nearly radial.

   As an aside, a more realistic model in the case of TPC pads with
fine segmentation would include the correlations
between track position measurements in nearby pads, which are mainly
due to knock-on electrons (i.e. delta-rays) spanning more than one
pad. However, because the momentum resolution is almost independent
of the pad segementation it will be assumed that a finely segmented
configuration can be considered to be equivalent to a configuration
with coarse enough sampling that the statistical independence
represented by equation~\ref{eq:sigmatpc} is still obeyed yet
equation~\ref{eq:nggfour} is also still valid.

   The statistical precision of a TPC's momentum resolution is limited
by the diffusion of the electrons as they drift through the TPC gas.
The diffusion
limit can be calculated from the ALEPH parameters~\cite{alephtpc}:
\begin{itemize}
  \item the specific ionization for a minimum ionising particle is
                                                   90 electrons/cm
  \item the magnetic field is 1.5 Tesla
  \item the transverse diffusion is $800\; {\rm \mu m.m}^{-1/2}$
                    per electron for a magnetic field of 1.5 Tesla
\end{itemize}
  For magnetic fields (B) of order a Tesla or larger, the diffusion
scales approximately as 1/B, so
the limiting precision on the track position, $\sigma_{limit}$ is:
\begin{eqnarray*}
\sigma_{limit} & = & 800\; \mu m \times (1.5\;{\rm Tesla/B}) \times
                                ({\rm drift\;length\;in\;metres})^{1/2}
                               / ({\rm spec.\: ion.})^{1/2}, \\
               & = & 67 {\rm \mu m.cm^{1/2}} \\
\end{eqnarray*}
for a magnetic field of 3 Tesla and a drift length of 2.5 m.

   As well as the statistical uncertainty due to electronic diffusion,
the TPC's momentum resolution will be degraded by systematic terms
in the point resolution. Examples of such effects are: imperfect
read-out (e.g. the ``angular pad effect''), space-charge effects
and possible systematics from the B field, E field and nearby tracks.

    A potentially large contribution to the resolution smearing comes from
the "angular pad effect": the avalanche electrons which are created at the
sense wires are displaced sideways due to the Lorentz effect. From the
information in the ALEPH references~\cite{alephtpc}
it follows easily that this gives an
uncertainty of $\sigma_{TPC} \simeq {\rm 70\; \mu m.cm^{1/2}}$ for a 1.5 Tesla
magnetic
field and the ALEPH geometry. Unfortunately, this term would be expected
to scale approximately in proportion to the magnetic field. For the ALEPH
read-out geometry it would be twice as large for a 3 Tesla field --
about $\sigma_{TPC} \simeq 140\; {\rm \mu m.cm^{1/2}}$. However,
one might assume that improvements can be made over the ALEPH
read-out, either within the sense-wire concept or by using a new type
of read-out such as microstrip gas chamber (MSGC) read-out.

    Solving equation~\ref{eq:resapprox} for $\sigma_{TPC}$ gives:
\begin{equation}
\sigma_{TPC} \simeq \frac{ \sigma(\frac{1}{p_t})
                           \times L^{2.5}
                           \times 0.3 \times B({\rm Tesla}) }{\sqrt{720}}.
                 \label{eq:tpcapprox}
\end{equation}
Substituting in the parameters assumed at the Munich workshop,
$B=3$ Tesla, $L=1.1$ m and $\sigma(\frac{1}{p_t})=2.3 \times 10^{-4}$,
gives
\begin{eqnarray*}
\sigma_{TPC} & = &  9.8 \times 10^{-6}\;{\rm m}^{3/2} \\
             & = &  98\; {\rm \mu m.cm^{1/2}}
                    \;\;\;\;\; {\rm (Munich\; TPC)}. \\
\end{eqnarray*}
For comparision, the ALEPH parameters,
$B=1.5$ Tesla, $L=1.32$ m and $\sigma(\frac{1}{p_t})=12 \times 10^{-4}$,
correspond to:
\begin{eqnarray*}
\sigma_{TPC} & = &  4.03 \times 10^{-5}\;{\rm m}^{3/2} \\
             & = &  403\; {\rm \mu m.cm^{1/2}}
                    \;\;\;\;\; {\rm (ALEPH\; TPC)}, \\
\end{eqnarray*}
i.e. a factor of four worse precision. However, it should be noted that only
48 percent of the radial coordinate of the ALEPH TPC is instrumented
for measuring the track position (21 pads of 3 cm length, in a total
radial span of 132 cm).
Since the average radial span of each of the ALEPH
pads is 6.3 cm, the predicted point resolution corresponds
to a pad resolution of $403/\sqrt{6.3}\;=\;160\;\mu {\rm m}$,
which agrees with the quoted ALEPH pad resolution of
170 microns.

   Very similar predictions for the point resolutions of both TPC's 
were obtained using the fitting procedure of section 3: 95 and
395 ${\rm \mu m.cm^{1/2}}$ for the Munich and ALEPH TPC's, respectively.
(These values were obtained assuming $N_{TPC}=40$; it was checked that
$N_{TPC}=20$ gave similar values.)

   Based on these results, it was considered reasonable to perform the
resolution studies on the CT for three different TPC point resolutions:
\begin{itemize}
   \item  $\sigma_{TPC}  =  95\; {\rm \mu m.cm^{1/2}}$
   \item  $\sigma_{TPC}  =  195\; {\rm \mu m.cm^{1/2}}$
   \item  $\sigma_{TPC}  =  390\; {\rm \mu m.cm^{1/2}}$.
\end{itemize}
The first and last of these values correspond roughly to the
momentum resolution assumed at Munich and to the ALEPH resolution,
respectively, and the middle value corresponds to an intermediate
resolution.

\section{Estimation of Subdetector Misalignments}

    For those momentum calculations which take misalignments into account,
the following gaussian misalignment uncertainties were assumed:
\begin{itemize}
  \item 5 microns for the vertex detector relative to beam spot
  \item 5 microns for the Si layer relative to vertex detector
  \item 80 microns for the TPC relative to Si layer.
\end{itemize}

   The 80 micron misalignment of the TPC was obtained using the
measured~\cite{alephtpc}
ALEPH transverse momentum ($p_t$) resolutions for the TPC alone, and
with the inner tracking chamber (ITC) and vertex detector (VD):
\begin{eqnarray*}
          \sigma_{p_t}/p_t^2 & = & 12.0 \times
                                10^{-4}\;({\rm GeV/c})^{-1}:
                         \hspace{0.8 cm}     {\rm TPC\;alone} \\
          \sigma_{p_t}/p_t^2 & = & 8.0 \times
                                10^{-4}\;({\rm GeV/c})^{-1}:
                         \hspace{0.8 cm}     {\rm TPC\;+\;ITC} \\
          \sigma_{p_t}/p_t^2 & = & 6.0 \times
                                10^{-4}\;({\rm GeV/c})^{-1}:
                         \hspace{0.8 cm}     {\rm TPC\;+\;ITC\;+\;VD}, \\
\end{eqnarray*}
as follows.

   As a very crude simulation of the ALEPH geometry, momentum
resolutions were obtained, using the method of section 3, with
the correct ALEPH TPC parameters,
$B=1.5$ Tesla, $L=1.32$ m and $\sigma(\frac{1}{p_t})=12 \times
10^{-4}$,
and with a vertex detector layer at a radius of 6.3 cm with a
point resolution of
8 microns. This configuration gives the correct resolution,
$\sigma_{TPC} = 6.0 \times 10^{-4}\;({\rm GeV/c})^{-1}$,
for a misalignment of 80 microns, while the resolution is too good if the
misalignment is left out:
$\sigma_{TPC} = 4.6 \times 10^{-4}\;({\rm GeV/c})^{-1}$.

   This determination of the ALEPH misalignment
is unrealistically simple because the ALEPH vertex detector actually
contains two layers of microstrips with point resolutions of 12 cm,
and the ITC has been ignored. However, the misalignment is much bigger
than the point resolution of the vertex detector, and it was also checked
that adding a second vertex layer didn't change the amount of
misalignment needed.

\section{Results}

    Tables 1 and 2 summarize the estimates for the CT momentum
resolution for different combinations of the following model variations:
\begin{itemize}
    \item the three different TPC coordinate resolutions listed in the
          preceding section.
    \item With and without some misalignment between the TPC and VD.
    \item With and without a vertex constraint.
    \item Momentum resolutions for the TPC alone, TPC plus VD, and
          the entire CT: TPC, VD and SI.
    \item Three different choices for the radius of the SI layer
          and inner radius of the CT.
\end{itemize}

   As a further study, it was found that
increasing the radius of the SI layer generally gave little
improvement to the momentum resolution. For example, the momentum
resolution for perfect alignment, no vertex constraint and a TPC
resolution of 195 ${\rm \mu m.cm^{1/2}}$ improves from 0.93
(in units of $10^{-4}\; {\rm (GeV/c)^{-1}}$) to
0.85 when the SI radius is increased from 32 cm to 42 cm, and
remains at 0.85 when the SI radius is further increased to 52 cm.

\begin{table}
\centering
\begin{tabular}{|rr|c|c|c|}
\hline
    TPC Align./ & res. &     TPC   &   TPC+VD  &  TPC+VD+SI \\
\hline
           0  / &  95  &     2.28   &    0.84   &     0.71 \\   
           0  / & 195  &     4.63   &    1.48   &     0.92 \\
           0  / & 390  &     9.30   &    2.61   &     1.43 \\
                &      &            &           &          \\
          80  / &  95  &     2.28   &    1.40   &     0.89 \\
          80  / & 195  &     4.63   &    1.71   &     1.21 \\
          80  / & 390  &     9.30   &    2.65   &     1.84 \\
\hline
\end{tabular}
\caption{
{\small \bf
   Fitted momentum resolutions for different
   alignment and resolution assumptions, and using different combinations
   of the central tracking subdetectors. The TPC misalignment is in
   units of microns and the resolution in units of ${\rm \mu m.cm^{1/2}}$.
   The momentum resolutions are in units of
   $10^{-4} \times ({\rm GeV/c})^{-1}$.
   Values are given for the TPC alone, TPC plus vertex detector and
   TPC plus vertex detector plus a silicon microstrip layer at a
   radius of 32 cm.
   The momentum resolutions are in units of
   $10^{-4} \times ({\rm GeV/c})^{-1}$.
   The values have statistical uncertainties of
   approximately 5 percent. No vertex constraint is used.
}}
\label{tab:momres no vtx}
\end{table}

\begin{table}
\centering
\begin{tabular}{|rr|c|c|c|}
\hline
    TPC Align./ & res. &     TPC   &   TPC+VD  &  TPC+VD+SI \\
\hline
           0  / &  95  &     0.79   &    0.70   &     0.55 \\   
           0  / & 195  &     1.57   &    1.05   &     0.74 \\
           0  / & 390  &     3.03   &    1.39   &     1.11 \\
                &      &            &           &          \\
          80  / &  95  &     2.04   &    1.16   &     0.68 \\
          80  / & 195  &     2.47   &    1.30   &     1.24 \\
          80  / & 390  &     3.60   &    1.74   &     1.78 \\
\hline
\end{tabular} 
\caption{
{\small \bf
 Momentum resolutions as in table 1, except that a vertex
constraint is used.
}}
\label{tab:momres with vtx}
\end{table}

\section{Summary and Conclusions}

   By comparing the various resolution estimates in tables 1 and 2,
the reader should
be able to form some sort of idea of the expected momentum resolution
of the CT under both optimistic and conservative assumptions about
the subdetector performances, and to assess the expected level of
improvement from adding the SI layer.

     In section 3,
the TPC-alone momentum resolution that was estimated at the
Munich workshop -- $2.3 \times 10^{-4}\; ({\rm GeV/c})^{-1}$ --
was found to correspond to approximately a factor of 4 improvement
in the TPC point resolution over the ALEPH performance. It is not
yet clear whether or not such a level of improvement is achievable.
However, the central tracker resolution seems to be rather robust
against degradations in the TPC point resolutions or, in a simple
model, misalignments of the TPC with respect to the vertex detector.

    The addition of a silicon microstrip layer just inside the
TPC would give increased robustness to the momentum resolution
by providing a stable and  precise coordinate measurement at a
radius approaching the mid-radius of the central tracker.
This should improve the CT performance in two ways:
\begin{itemize}
  \item The precise coordinate information near the mid-point
should improve the statistical precision of the momentum fit.
   \item The alignment of the TPC with respect to the vertex detector
should improve with the addition of a stable time-independent
coordinate measurement near the TPC inner radius.
\end{itemize}
The first of these items can be seen from the tabulated results,
while the second is more difficult to quantify.

   Perhaps surprisingly, increasing the SI radius seems to give little
improvement to the momentum resolution. It seems that the improved
lever-arm of the Si is almost compensated for by the reduced resolution
of the TPC.

    In conclusion, it appears likely that the TPC-based central tracker
design which was discussed at the ECFA/DESY Munich workshop, including
a layer of silicon microstrips, could achieve a
momentum resolution of approximately $10^{-4}\; ({\rm GeV/c})^{-1}$.

\vskip 0.5 cm
 
\centerline{\large\bf Acknowledgements}
\vskip 0.5 cm

   This work was carried out at the DESY laboratory as part of the
collaborative effort to design a detector for the conceptual design
report of the ECFA/DESY 500 GeV electron-positron linear collider.
Discussions with Ron Settles have been very helpful for understanding
some of the issues involved in TPC design and performance.
The author would particularly like to thank the members of the
Hamburg OPAL group for their help and support.

\pagebreak

\end{document}